\documentclass[12pt]{article}

\usepackage{amstext,amsmath,amssymb,amsfonts,bbm,slashed}
\usepackage[latin1]{inputenc}
\usepackage{epsfig}
\usepackage{hyperref}
\usepackage{amsthm}
\usepackage{subfigure}
\usepackage{color}
\usepackage{multirow}

\newtheorem{lemma}{Lemma}

\newtheorem{theorem}{Theorem}
\newtheorem{proposition}{Proposition}

\newcommand{\bea}{\begin{eqnarray}}
\newcommand{\eea}{\end{eqnarray}}
\newcommand{\beq}{\begin{equation}}
\newcommand{\eeq}{\end{equation}}

\newcommand{\bpro}{\begin{pro}}
	\newcommand{\epro}{\end{pro}}
\newcommand{\blem}{\begin{lem}}
	\newcommand{\elem}{\end{lem}}
\newcommand{\bdfn}{\begin{dfn}}
	\newcommand{\edfn}{\end{dfn}}
\newcommand{\bcor}{\begin{cor}}
	\newcommand{\ecor}{\end{cor}}
\newcommand{\bthm}{\begin{thm}}
	\newcommand{\ethm}{\end{thm}}
\newcommand{\bex}{\begin{ex}}
	\newcommand{\eex}{\end{ex}}
\newcommand{\brmk}{\begin{rmk}}
	\newcommand{\ermk}{\end{rmk}}
\newcommand{\bpr}{\begin{pr}}
	\newcommand{\epr}{\end{pr}}

\makeatletter

\begin{document}

	\begin{center}
		
		{\LARGE\bf Hamiltonian Dynamics of  a spaceship in Alcubierre and G\"{o}del metrics: \\ Recursion operators and underlying master symmetries\\}

		\vspace{15pt}
		
		{\large   Mahouton Norbert Hounkonnou $^{a,b} $ , Mahougnon Justin Landalidji $^{a,b}, $ and \\ Melanija Mitrovi\'c $^{a,b,c} $
		}
		
		\vspace{15pt}
		
		{  {\small $a$)\,\,
			International Chair of Mathematical Physics
			and Applications  (ICMPA-UNESCO Chair)\\
			University of Abomey-Calavi, 072 B.P. 50 Cotonou, Republic of Benin\\ 
			$b$) Centre International de Recherches et d'Etude Avanc\'{e}es en Sciences Math\'{e}matiques \& Informatiques et Applications (CIREASMIA), 072 B.P. 50 Cotonou, Republic of Benin
			E-mails: { norbert.hounkonnou@cipma.uac.bj with copy to
				hounkonnou@yahoo.fr} \\
			E-mails: { landalidjijustin@yahoo.fr}   \\
			\vspace{5pt}
			$c$)\,\, Center of Applied Mathematics of the Faculty of Mechanical Engineering (CAM-FMEN),
			University of Ni\v s,  Serbia \\
			E-mails: {  melanija.mitrovic@masfak.ni.ac.rs  }}
		}

	\end{center}
	\vspace{10pt}
	\begin{abstract}
	We study the Hamiltonian dynamics of a   spaceship  in the background of Alcubierre and G\"{o}del metrics. We derive the Hamiltonian vector fields governing the system evolution, construct and discuss related recursion operators generating the constants of motion. Besides, we characterize relevant master symmetries.
		
		
	\end{abstract}
\section{Introduction}\label{sec1}

In 1949, G\"{o}del \cite{go} found a solution of the Einstein equations
corresponding to a homogeneous mass distribution that
rotates at each point of the space \cite{gra}. This  distribution of
matter causes unusual effects, such as the existence of closed timelike curves (CTCs).  
However,  G\"{o}del's
metric has the advantage that it is  rather of compact form and most calculations can be carried out analytically \cite{ka}. 
The  G\"{o}del solution makes it apparent that
general relativity permits solutions with closed time-like worldlines, even when
the metric possesses a local Lorentzian character that ensures an inherited
regular chronology, and therefore, the local validation of the causality principle \cite{kli}.
In recent decades, the study of CTCs  attracted the attention of several authors (see $e.g.$, \cite{ahm1,ahm2,gra,ka,kli,sa}). 
In particular, in 2004, Kajari {\it et al.} presented exact expressions for the Sagnac effect of G\"{o}del Universe. 
In their work,  
they proposed a formulation of the Sagnac time delay in terms of invariant
physical quantities and showed that this result is very close to the analogous formula of
the Sagnac time delay of a rotating coordinate system in Minkowski spacetime. 

Moreover, it is known that in general relativity, faster-than-light (FTL) speed is only forbidden locally \cite{gra}. This is not as exotic as
it might seem at first glance.  For instance, the expansion of
the Universe can make that two distant galaxies move at
FTL speed between them, while each one is moving locally
inside its light cone. The opposite might be possible too: if
the spacetime were contracting fast enough, each galaxy
were moving near the speed of light locally (inside its light
cone) in opposite directions, but globally both were getting
closer. With these considerations in mind, Alcubierre  introduced in 1994 \cite{alc} the so-called warp drive
metric (WDM), within the framework of general relativity, which allows in principle for superluminal motion,
that is, FTL travel \cite{ga,gra}. This Alcubierre's idea
consists to create in the front of an object (a spaceship for example), a spacetime
contraction, and in the back, a spacetime dilation. Thus, the
contraction will pull the object forward, and the dilation
will push the object forward too \cite{gra}. Locally, the object will be
inside its light cone, but due to this spacetime manipulation, it would move FTL as compared with $c$, the speed of light
in flat-spacetime vacuum. The  objet is
within the so-called warp bubble. In this way, the objet can travel at arbitrarily high speeds, without violating the laws of special and general relativity, or other known physical laws \cite{ga}. Following Alcubierre's idea, many investigations were done (see $e.g.$ \cite{ga,gra}).

In addition, in the last few decades,  there was a renewed interest in
completely integrable Hamiltonian systems
(IHS), the concept of which goes back to Liouville in 1897 \cite{lio}
and Poincar\'{e} in 1899 \cite{pau}. In short, IHS are defined as nonlinear differential equations admitting a Hamiltonian description and possessing sufficiently many constants of motion so that they can be integrated by quadratures \cite{fil1}. 
Many of these systems obey Hamiltonian dynamics with respect to two compatible symplectic structures  \cite{mag1}, \cite{gel}, \cite{vil1}, permitting a geometrical interpretation of the so-called recursion operator \cite{lax}.  A description of integrability
working both for systems with finitely many degrees of freedom and for field theory
can be given in terms of invariant, diagonalizable mixed $(1,1)$-tensor field, having bidimensional eigenspaces and vanishing Nijenhuis torsion. One of powerful methods of describing IHS with involutive Hamiltonian functions or constants of motion uses the recursion operator admitting a vanishing Nijenhuis torsion. 
In 2015,  Takeuchi  constructed  recursion operators of Hamiltonian vector fields of geodesic flows for some  Riemannian and Minkowski  metrics \cite{ta2}, and  obtained related constants of motion. 
In his work, he  used five particular solutions of the Einstein equation in the Schwarzschild,  Reissner-Nordström,   Kerr,  Kerr-Newman, and  FLRW metrics, and  constructed recursion operators  inducing the complete integrability of the Hamiltonian functions. In 2019, we  investigated  the same problem in a noncommutative Minkowski phase space  and  the Kepler problem in a deformed phase space, and obtained associated constants of motion, \cite{hkn1}, \cite{hkn2}. 

Since the work by Magri, the integrability associated with bi-Hamiltonian structures  \cite{mag1} 
became one of the most efficient methods used for the integrability of evolution
equations in both finite and infinite dimensional dynamical systems \cite{lax,smir}.
When a completely
integrable Hamiltonian system   admits a bi-Hamiltonian construction,
one can generate infinite hierarchies of conserved quantities using the construction by Oevel \cite{oev} based on scaling invariances and master
symmetries \cite{fer2,smir2}.
In 1997 and 1999, Smirnov \cite{smir,smir2} formulated a constructive
method of transforming a completely integrable Hamiltonian system,
in Liouville's sense, into Magri-Morosi-Gel'fand-Dorfman's (MMGD)
bi-Hamiltonian form. 
In 2005,  Ra\~{n}ada  \cite{ra}
proved the existence of a bi-Hamiltonian structure arising from a
non-symplectic symmetry  as well as the existence of master
symmetries and additional integrals of motion (weak
superintegrability) for certain particular cases. Recently, in 2021 \cite{hkn3}, we  also constructed a hierarchy of bi-Hamiltonian
structures for the Kepler problem,  and   computed conserved quantities using  related master
symmetries. 

In the present  work, we address the Hamiltonian dynamics {of  a spaceship} in Alcubierre and G\"{o}del metrics.
We derive  related recursion operators and discuss their relevant master symmetries. We prove that the two models satisfy the same dynamics and exhibit a set of similar  master symmetries.

The paper is organized as follows.   In Section 2, we give the main tools  used in this work. In Section
\ref{sec3}, we give the
Hamiltonian function, the symplectic form and the vector field describing
the Hamiltonian dynamics of  a spaceship  in the Alcubierre metric and construct the associated recursion operators. In Section \ref{sec4}, we perform the same study,  as in Section \ref{sec3},
in the G\"{o}del  metric. 
In Section \ref{sec5}, we
introduce bi-Hamiltonian structures, define the hierarchy of master
symmetries and compute the corresponding conserved quantities. In Section
\ref{sec6}, we end with some concluding remarks.
\section{Recursion operator and master symmety} \label{sec2}
A  characterization of integrable Hamiltonian systems is given by De Filippo {\it et al.} through  the following Theorem \cite{fil1}:
\begin{theorem}\label{thi}
	Let $X$ be a dynamical vector field on a $2n$-dimensional manifold $\mathcal{M}$. If the vector field $X$ admits a diagonalizable mixed $(1, 1)$-tensor field $T$ which is invariant under $X,$ has a vanishing Nijenhuis torsion and  doubly degenerate eigenvalues
	with nowhere vanishing differentials, then there exists a symplectic structure and a
	Hamiltonian function $H$ such that the vector field $X$ is a separable Hamiltonian vector
	field of $H,$ and $H$ is completely integrable with respect to the symplectic structure.
\end{theorem}
Such a $(1, 1)$-tensor field $T$ is called a recursion operator of $X$.
In the particular case of $\mathbb{R}^{2n},$ a recursion operator can be constructed as follows \cite{ta2}:
\begin{lemma} \label{lem_1} \
	Let us consider vector fields
	{\small\[
		X_{l} = - \dfrac{\partial}{\partial{x_{n+l}}}, \ \ l = 1,...,n
		\]}
	on $\mathbb{R}^{2n}$ and let $T$ be a $(1,1)$-tensor field on $\mathbb{R}^{2n}$ given by
	{\small\[
		T = \sum_{i =1}^{n}x_{i}\Bigg(\dfrac{\partial}{\partial{x_{i}}} \otimes dx_{i} + \dfrac{\partial}{\partial{x_{n+i}}} \otimes dx_{n+i}\Bigg).
		\]}
	Then, we have that the Nijenhuis torsion $ \mathcal{N}_{T}$ and the Lie derivative $\mathcal{L}_{X_{l}}$ of {$T$} are vanishing,  $i.e.,$ 
	{\small\begin{eqnarray*} (\mathcal{N}_{T})^{h}_{ij}:=T^{k}_{i}\dfrac{\partial{T^{h}_{j}}}{\partial{x^{k}}}  - T^{k}_{j}\dfrac{\partial{T^{h}_{i}}}{\partial{x^{k}}} + T^{h}_{k}\dfrac{\partial{T^{k}_{i}}}{\partial{x^{j}}}  - T^{h}_{k}\dfrac{\partial{T^{k}_{j}}}{\partial{x^{i}}} = 0, \ \mbox{and} \ 
			\mathcal{L}_{X_{l}}T = 0,
	\end{eqnarray*}}	
	$i.e.,$   the $(1,1)$-tensor field $T$ is a recursion operator of $X_{l}, (l = 1,...,n).$
\end{lemma}
Given a general dynamical system defined on a $2n$-dimensional manifold $\mathcal{Q}$ \cite{smir},
\begin{equation} \label{dyn}
\dot{x}(t) = X(x), \quad x \in \mathcal{Q}, \quad X \in \mathcal{T}\mathcal{Q},
\end{equation}
where $\mathcal{T}\mathcal{Q}$
is the tangent bundle of $\mathcal{Q}$.

If this system \eqref{dyn} admits two different Hamiltonian representations:
\begin{equation} \label{dyn2}
\dot{x}(t) = X_{H_{1},H_{2}} = \mathcal{P}_{1}dH_{1} = \mathcal{P}_{2}dH_{2}, 
\end{equation}
its integrability as well as many other properties are subject to Magri's approach,
$i.e.,$  the bi-Hamiltonian vector field $X_{H_{1},H_{2}}$ is defined by two pairs of Poisson bivectors $\mathcal{P}_{1}, \mathcal{P}_{2}$ and Hamiltonian functions $ H_{1},H_{2}.$ 
$\mathcal{P}_{1}$ and $\mathcal{P}_{2}$ are  compatible Poisson bivectors with vanishing Schouten-Nijenhuis bracket \cite{du}:
$
[\mathcal{P}_{1}, \mathcal{P}_{2}]_{NS} = 0.
$ \\
Such a manifold $\mathcal{Q}$ equipped with two Poisson bivectors is called a  double Poisson manifold and the quadruple $(\mathcal{Q},\mathcal{P}_{1},\mathcal{P}_{2},X_{H_{1},H_{2}} )$ is called a bi-Hamiltonian system.

In differential geometric terms, a vector field   $Y$ on the cotangent bundle  $\mathcal{T}^{\ast}\mathcal{Q}$ that satisfies
{\small\begin{equation*}
	[X_{H'}, Y] \neq 0 , \quad [X_{H'}, X] = 0, \quad [X_{H'}, Y]  = X,
	\end{equation*}}
is called a master symmetry or a generator of symmetries of degree  $m = 1$ for the Hamiltonian vector field $X_{H'}$ \cite{cas,dam,fer2,ra,ra3}.

\section{Recursion operator of a Hamiltonian vector field in the Alcubierre metric} \label{sec3}

In this work, without loss of generality, we consider the following particular Alcubierre metric \cite{alc}: 
{\small\begin{equation*} \label{metr1}
	ds^{2} = - dt^{2} + (dx - v_{s}f(r_{s})dt)^{2} + dy^{2} + dz^{2}
	\end{equation*}}
describing the motion of a spaceship  along the  $x$-axis of a cartesian coordinate system such as: 
{\small\begin{align*}
	& \alpha = 1, \ \beta_{2} = -  v_{s}f(r_{s}), \ \beta_{3} = \beta_{4} = 0 , \ \gamma_{ij} = \delta_{ij}, \ \mbox{( $\delta_{ij}$  is the Kronecker symbol   )}, \\
	& v_{s} = \dfrac{dx_{s}(t)}{dt} , \ r_{s}(t) = ((x - x_{s}(t))^{2} + y^{2} + z^{2})^{1/2}  \\
	& f(r_{s}) = \dfrac{\tanh(\sigma(r_{s} + R)) - \tanh(\sigma(r_{s} - R))}{2\tanh(\sigma R)}, 
	\end{align*}}
where $\sigma > 0$, $R > 0$ are arbitrary parameters, and 
{\small\begin{equation*}
	\lim\limits_{ \sigma \rightarrow \infty} f(r_{s}) = \left\{
	\begin{array}{ll}
	1 \quad \mbox{for} \ \ r_{s} \in ]- R, R[   \\ 
	\frac{1}{2}  \quad \mbox{for} \ \ r_{s} \in \{- R, R\}\\
	0 \quad \mbox{otherwise}.
	\end{array}
	\right.
	\end{equation*}}

Under the limit $ \sigma \rightarrow \infty,$ with $r_{s} \in ]- R, R[,$ this particular Alcubierre metric  becomes 
{\small\begin{equation} \label{metr2}
	ds^{2} = - dt^{2} + (dx - v_{s}dt)^{2} + dy^{2} + dz^{2},
	\end{equation}}
where the tensor metric and its inverse are given by
{\small\begin{equation*} \label{teng}
	g_{\nu\mu} = \left(
	\begin{array}{cccc}
	- (1 - v_{s}^{2}) & - v_{s}& 0 & 0\\
	- v_{s} & 1 & 0 & 0 \\
	0 & 0 & 1 & 0\\
	0 & 0 & 0  & 1 \\
	\end{array}
	\right), \quad \mbox{and} \quad g^{\nu\mu} = \left(
	\begin{array}{cccc}
	- 1  & - v_{s}& 0 & 0\\
	- v_{s} & (1 - v_{s}^{2}) & 0 & 0 \\
	0 & 0 & 1 & 0\\
	0 & 0 & 0  & 1 \\
	\end{array}
	\right).
	\end{equation*}}
Geometrically, this spacetime can be interpreted as follows \cite{alc}:\\
Firstly, since $\gamma_{ij} = \delta_{ij}$, the $3$-geometry of the hypersurfaces is then always flat. Secondly, the fact that the lapse is given by $ \alpha = 1$ implies that the time-like curves normal to these hypersurfaces are geodesics, $i.e.,$ the Eulerian observers are in free fall. Notice that spacetime, however,
is not flat due to the presence of a non-uniform shift. Finally, since the shift vector vanishes for $ r_{s} \gg R$,  at any time $t$, spacetime will be essentially flat everywhere
except within a region with a radius of order $R$, centred at the point $(x_{s}(t),0,0)$.

Let now
$\mathcal{Q} = \mathbb{R}^{4} := \{ q^{1} = t, q^{2} = x, q^{3} = y,  q^{4} = z \} $ be  the manifold  describing the configuration space 
and $\mathcal{T}^{\ast}\mathcal{Q} = \mathcal{Q} \times \mathbb{R}^{4}$ be
the cotangent bundle with the local coordinates $(q,p),$ and
the natural symplectic
structure $\omega_{A} :
\mathcal{T}\mathcal{Q} \longrightarrow \mathcal{T}^{\ast}\mathcal{Q}$  given by
{\small\begin{equation*}\label{Ksy}
	\omega_{A} = \sum_{\nu = 1}^{4} dp_{\nu} \wedge dq^{\nu},
	\end{equation*} }
where $\mathcal{T}\mathcal{Q}$
is the tangent bundle.
By definition, $\omega_{A}$ is
non-degenerate. It induces the map $\mathcal{P}_{A}$:
$\mathcal{T}^{\ast}\mathcal{Q} \longrightarrow \mathcal{T}\mathcal{Q}$,  called  bivector field,
defined by
{\small\begin{equation*}\label{Kbi2}
	\mathcal{P}_{A} = \sum_{\nu = 1}^{4} \dfrac{\partial}{\partial p_{\nu} }\wedge
	\dfrac{\partial}{\partial q^{\nu}},
	\end{equation*}}  which is  the inverse map of $\omega_{A}, $ $i.e.,$   $ \omega_{A} \circ \mathcal{P}_{A} = \mathcal{P}_{A} \circ \omega_{A} = 1$
\cite{vil2}. In this case, the Hamiltonian vector field $X_{f}$
of a Hamiltonian function $f$ is given by
{\small\begin{equation*}\label{Kvec}
	X_{f} = \mathcal{P}_{A}df.
	\end{equation*} }
On the  cotangent bundle  $\mathcal{T}^{\ast}\mathcal{Q},$  \eqref{metr2} takes the form:

\begin{equation*} \label{metric3}
ds^{2} = - (dq^{1})^{2} + (dq^{2} - v_{s}dq^{1})^{2} + (dq^{3})^{2} + (dq^{4})^{2}.
\end{equation*}

In our framework, the  Hamiltonian function  $\mathcal{H}_{A}$ describing the dynamics {of  a spaceship} in the Alcubierre metric and 
its corresponding $1$-form  $d\mathcal{H}_{A} \in \mathcal{T}^{\ast}\mathcal{Q}$
are given by:
\begin{equation} \label{Hama}
\mathcal{H}_{A} :=\dfrac{1}{2} \sum_{\nu, \mu = 1}^{4} g^{\nu \mu}p_{\nu}p_{\mu} = \dfrac{1}{2}\Bigg( -p_{1}^{2} - v_{s}p_{1}p_{2} + (1 -  v_{s}^{2})p_{2}^{2} + p_{3}^{2} + p_{4}^{2} \Bigg),
\end{equation}
and 
\begin{align*} \label{dh}
d\mathcal{H}_{A}  &= - ( p_{1} + v_{s}p_{2}) dp_{1} + (-v_{s}p_{1} + (1 -  v_{s}^{2})p_{2} )dp_{2} \cr 
& + p_{3}dp_{3} + p_{4}dp_{4} - \dot{v}_{s}(p_{1} + v_{s}p_{2} )p_{2}dq^{1},
\end{align*}
respectively.
Then, the  Hamiltonian vector field of $\mathcal{H}_{A}$ with respect to the symplectic structure $\omega_{A}$ is derived as
\begin{align*}
X_{\mathcal{H}_{A}} := \{\mathcal{H}_{A}, .\} & = - ( p_{1} + v_{s}p_{2})\dfrac{\partial}{\partial q^{1}} + (-v_{s}p_{1} + (1 -  v_{s}^{2})p_{2} )\dfrac{\partial}{\partial q^{2}} \cr & + p_{3}\dfrac{\partial}{\partial q^{3}}  + p_{4}\dfrac{\partial}{\partial q^{4}} + \dot{v}_{s}(p_{1} + v_{s}p_{2} )p_{2} \dfrac{\partial}{\partial p_{1}}. 
\end{align*}

This Hamiltonian vector field satisfies the required condition for a Hamiltonian system, $i.e.,$
\begin{equation*}
\iota_{_{X_{\mathcal{H}_{A}}}} \omega_{A} = -d\mathcal{H}_{A},
\end{equation*}
where $\iota_{_{X_{\mathcal{H}_{A}}}} \omega_{A}$ is  the interior product of $ \omega_{A} $ with respect to the Hamiltonian vector field $X_{\mathcal{H}_{A}}.$  Hence, the triplet $(\mathcal{T}^{\ast}\mathcal{Q},\omega_{A} , \mathcal{H}_{A})$ is a Hamiltonian system.

In the sequel, we consider the Hamilton-Jacobi equation with respect to the Hamiltonian function \eqref{Hama} and  introduce a generating function $W$ satisfying the following canonical transformations \cite{abr,arn}  :
\begin{equation*}\label{can}
p = \dfrac{\partial{W}}{\partial{q}} \quad \mbox{and} \quad P = - \dfrac{\partial{W}}{\partial{Q}}. 
\end{equation*}
Since the Hamiltonian function $\mathcal{H}_{A} $ does not explicitly depend on the time $t,$ then, setting $V = W - Et,$ it is possible to find an additive separable solution:
\begin{equation*}
W = W_{1}(q^{1}) + W_{2}(q^{2}) + W_{3}(q^{3}) + W_{4}(q^{4}). 
\end{equation*} 
The Hamilton-Jacobi equation 
\cite{arn}
\begin{eqnarray*} \label{Eq_3_17}
	\dfrac{\partial V}{\partial t} + \mathcal{H}_{A}\bigg(\dfrac{\partial V}{\partial q} \bigg/ q/t \bigg) = 0
\end{eqnarray*}
is then reduced to the nonlinear equation
\begin{equation} \label{HJE}
E  = \dfrac{1}{2}\Bigg\{-\Bigg(\dfrac{\partial{W}}{\partial{q^{1}}}\Bigg)^{2} - 2v_{s}\dfrac{\partial{W}}{\partial{q^{1}}}\dfrac{\partial{W}}{\partial{q^{2}}} + (1 - v_{s}^{2})\Bigg(\dfrac{\partial{W}}{\partial{q^{2}}}\Bigg)^{2} + \Bigg(\dfrac{\partial{W}}{\partial{q^{3}}}\Bigg)^{2}  + \Bigg(\dfrac{\partial{W}}{\partial{q^{4}}}\Bigg)^{2}\Bigg\},
\end{equation}
where $E$ is a constant.
We  notice  that the Hamiltonian function does not include $ q^{2}, q^{3},$ and $q^{4}$. Then, putting
\begin{equation*} \label{HJE1}
\dfrac{dW_{2}}{dq^{2}} (q^{2}) = \alpha_{0}, \  \dfrac{dW_{3}}{dq^{3}} (q^{3}) = \beta_{0}, \  \dfrac{dW_{4}}{dq^{4}} (q^{4}) = \gamma_{0},
\end{equation*} 
where $\alpha_{0}, \beta_{0},$  $\gamma_{0}$ are constants and $2E \leq  ( \beta_{0}^{2} + \gamma_{0}^{2} + \alpha_{0}^{2}) $, \eqref{HJE} becomes

\begin{equation*} \label{HJE2}
\Bigg(\dfrac{dW_{1}}{dq^{1}}\Bigg)^{2} + 2\alpha_{0} v_{s}\dfrac{dW_{1}}{dq^{1}} + K = 0, 
\end{equation*} 
where $K = 2E - (\beta_{0}^{2} + \gamma_{0}^{2} + (1 - v_{s}^{2})\alpha_{0}^{2}).$

Now, setting $ \Psi = \dfrac{dW_{1}}{dq^{1}}$, with $W_{1} (0) = 0,$ we obtain the following quadratic equation
\begin{equation*} \label{HJE3}
\Psi^{2} + 2\alpha_{0} v_{s}\Psi + K = 0, 
\end{equation*}
with   $\Delta_{A} = - 8E + 4( \beta_{0}^{2} + \gamma_{0}^{2} + \alpha_{0}^{2})  \geq 0$. Then, we have two possible cases: \\ $\Delta_{A} > 0$ or $\Delta_{A} = 0.$

\begin{itemize}
	\item[(i)] For $ \Delta_{A} > 0$, we get 
	{\small \begin{equation*} \label{HJE5}
		\Psi_{1} = - v_{s}\alpha_{0} + \sqrt{( \beta_{0}^{2} + \gamma_{0}^{2} + \alpha_{0}^{2}) - 2E}  , \quad \mbox{
			and} \ \    \Psi_{2} = - (v_{s}\alpha_{0} + \sqrt{( \beta_{0}^{2} + \gamma_{0}^{2} + \alpha_{0}^{2}) - 2E}) 
		\end{equation*} }
	leading to solutions for the generating function $W:$
	\begin{align*}
	& W_{a} = - \alpha_{0} q_{s} - (\sqrt{( \beta^{2} + \gamma_{0}^{2} + \alpha_{0}^{2}) - 2E}) q^{1} + \alpha_{0} q^{2} + \beta_{0} q^{3} + \gamma_{0} q^{4}, \\
	& W_{b} = - \alpha_{0} q_{s} + (\sqrt{( \beta_{0}^{2} + \gamma_{0}^{2} + \alpha_{0}^{2}) - 2E}) q^{1} + \alpha_{0} q^{2} + \beta_{0} q^{3} + \gamma_{0} q^{4},
	\end{align*}
	
	which, in terms of $q^{i}$ and $ Q^{i},$ are expressed as:
	\begin{align} 
	& W_{a} = -Q^{2}q_{s} - \Bigg(\sqrt{\sum_{k=2}^{4} (Q^{k})^{2} - 2Q^{1}}  - v_{s}Q^{2} \Bigg)q^{1} + \sum_{k=2}^{4} Q^{k}q^{k},\label{s1} \\
	& W_{b} = -Q^{2}q_{s} + \Bigg(\sqrt{\sum_{k=2}^{4} (Q^{k})^{2} - 2Q^{1}}  - v_{s}Q^{2} \Bigg)q^{1} + \sum_{k=2}^{4} Q^{k}q^{k} \label{s2},
	\end{align}
	where $Q^{1}=E, Q^{2} = \alpha_{0}, Q^{3} = \beta_{0},  Q^{4} = \gamma_{0},$  and  $\bigg(\displaystyle\sum_{k=2}^{4} (Q^{k})^{2} - 2Q^{1}\bigg) > 0.$ 
	
	In the following, we  consider  each of these solutions to derive the relationship between the canonical coordinate systems $(Q,P)$ and $(q,p)$. \\
	$\bullet$ For $W = W_{a},$  we obtain the relations:
	\begin{equation}\label{Eq_3_21}
	\left\{
	\begin{array}{ll}
	p_{1}= - \displaystyle\sqrt{\sum_{k=2}^{4} (Q^{k})^{2} - 2Q^{1}}  - v_{s}Q^{2} \\
	p_{2} =  Q^{2}     \\
	p_{3} =  Q^{3}  \\
	p_{4} =  Q^{4} 
	\end{array}
	\right.  ; \quad \left\{
	\begin{array}{ll}
	q^{1} = - P_{1} \displaystyle\sqrt{\sum_{k=2}^{4} (Q^{k})^{2} - 2Q^{1}}      \\
	q^{2} = q_{s} - P_{2} -  Q^{2}P_{1}   \\
	q^{3} =  - P_{3} -  Q^{3}P_{1} \\
	q^{4} =  - P_{4} -  Q^{4}P_{1} 
	\end{array}
	\right.
	\end{equation}
	
	\begin{equation}\label{Eq_3_21_2}
	\left\{
	\begin{array}{ll}
	P_{1} = \dfrac{q^{1}}{ p_{1} + v_{s}p_{2}} \\
	P_{2} = - \dfrac{p_{2}q^{1}}{ p_{1} + v_{s}p_{2}} + q_{s} - q^{2}\\
	P_{3} = - \dfrac{p_{3}q^{1}}{ p_{1} + v_{s}p_{2}}  - q^{3} \\
	P_{4} = - \dfrac{p_{4}q^{1}}{ p_{1} + v_{s}p_{2}}  - q^{4}
	\end{array}
	\right. ; \quad \left\{
	\begin{array}{ll}
	Q^{1} = \mathcal{H} \\
	Q^{2} =  p_{2} \\
	Q^{3} =  p_{3} \\
	Q^{4} =  p_{4}.
	\end{array}
	\right.
	\end{equation}
	$\bullet$ For $W = W_{b},$ we have:
	\begin{equation}\label{Eq_s1_2}
	\left\{
	\begin{array}{ll}
	p_{1}= \displaystyle\sqrt{\sum_{k=2}^{4} (Q^{k})^{2} - 2Q^{1}}  - v_{s}Q^{2} \\
	p_{2} =  Q^{2}     \\
	p_{3} =  Q^{3}  \\
	p_{4} =  Q^{4} 
	\end{array}
	\right.  ; \quad \left\{
	\begin{array}{ll}
	q^{1} =P_{1} \displaystyle\sqrt{\sum_{k=2}^{4} (Q^{k})^{2} - 2Q^{1}}      \\
	q^{2} = q_{s} - P_{2} -  Q^{2}P_{1}   \\
	q^{3} =  - P_{3} -  Q^{3}P_{1} \\
	q^{4} =  - P_{4} -  Q^{4}P_{1} 
	\end{array}
	\right.
	\end{equation}
	\begin{equation}\label{Eq_s_1_2_2}
	\left\{
	\begin{array}{ll}
	P_{1} = \dfrac{q^{1}}{ p_{1} + v_{s}p_{2}} \\
	P_{2} = - \dfrac{p_{2}q^{1}}{ p_{1} + v_{s}p_{2}} + q_{s} - q^{2}\\
	P_{3} = - \dfrac{p_{3}q^{1}}{ p_{1} + v_{s}p_{2}}  - q^{3} \\
	P_{4} = - \dfrac{p_{4}q^{1}}{ p_{1} + v_{s}p_{2}}  - q^{4}
	\end{array}
	\right. ; \quad \left\{
	\begin{array}{ll}
	Q^{1} = \mathcal{H}_{A} \\
	Q^{2} =  p_{2} \\
	Q^{3} =  p_{3} \\
	Q^{4} =  p_{4}.
	\end{array}
	\right.
	\end{equation}
	
	Defined in the coordinate system $(Q,P)$, the \textit{Alcubierre} symplectic form and the vector field are given, respectively,  as:
	\begin{equation*} \label{Eq_3_23}
	\omega_{A} = \sum_{\nu=1}^{4} dP_{\nu}\wedge dQ^{\nu},\ X_{\mathcal{H}_{A}} := \{\mathcal{H}_{A},.\} = -\dfrac{\partial}{\partial{P_{1}}}.
	\end{equation*}

	In this condition, a tensor field $T_{A}$ of $(1,1)$-type can be expressed as:
	\begin{equation*} \label{Eq_tensor}
	T_{A} = \sum_{\nu=1}^{4}Q^{\nu}\Bigg(\dfrac{\partial}{\partial{P_{\nu}}}\otimes dP_{\nu} + \dfrac{\partial}{\partial{Q^{\nu}}}\otimes dQ^{\nu}\Bigg).
	\end{equation*}
	Taking $x_{\nu}= Q^{\nu}$ and $x_{\nu +4} = P_{\nu}$ in Lemma \ref{lem_1}, where $\nu = 1,2,3,4,$ the tensor field $T_{A} $ takes the  form :
	\[
	T_{A} =\sum_{\nu=1}^{4}Q^{\nu}\Bigg(\dfrac{\partial}{\partial{P_{\nu}}}\otimes dP_{\nu} + \dfrac{\partial}{\partial{Q^{\nu}}}\otimes dQ^{\nu}\Bigg) = \sum_{i,j =1}^{2n}(T_{A})^{i}_{j}\dfrac{\partial}{\partial{x^{i}}}\otimes dx^{j},
	\]
	with $x\equiv (Q^{1},...,Q^{4}, P_{1},...,P_{4}).$ The matrix $(T_{A})^{i}_{j} $ is given by
	\[
	(T_{A})^{i}_{j} = \left(
	\begin{array}{cc}
	^{t}G & O \\
	O & G \\
	\end{array}
	\right), \ \ \
	G =\left(
	\begin{array}{cccc}
	Q^{1} & 0 & 0 & 0 \\
	0 & Q^{2} & 0 & 0 \\
	0 & 0 & Q^{3} & 0 \\
	0 & 0 & 0 & Q^{4} \\
	\end{array}
	\right).
	\]
	The tensor $T_{A}$ satisfies $\mathcal{L}_{X_{\mathcal{H}_{A}}}T_{A} = 0$, $\mathcal{N}_{T_{A}} = 0$ and $deg Q^{\nu} = 2$ proving that $T_{A}$ is a recursion operator of $X_{\mathcal{H}_{A}}$. The constants of motion  are :
	\[
	Tr(T_{A}^{h}) = 2((Q^{1})^{h} + (Q^{2})^{h} + (Q^{3})^{h} + (Q^{4})^{h}), \quad \ h \in \mathbb{N}.
	\]
	Reverting back to the  original coordinate system $(q,p),$ the generating functions $W_{a}$ and $W_{b}$  lead to the following result for the Alcubierre metric $ds^{2} = - (dq^{1})^{2} + (dq^{2} - v_{s}dq^{1})^{2} + (dq^{3})^{2} + (dq^{4})^{2}:$
	\begin{proposition}
		\label{Pro_3_1} \ Provided the conditions
		\begin{enumerate}
			\item [(1)]   $ \dfrac{\dot{v}_{s}}{v_{s}} = -\dfrac{1}{q^{1}} ;$
			\item [(2)]  $\dfrac{\ddot{v}_{s}}{\dot{v}_{s}}v^{h - 1}_{s} = \bigg(\dfrac{p_{1}}{p_{2}}\bigg)^{h - 1} (p_{1} + v_{s}p_{2})^{h - 1}, \quad h \in \mathbb{N}, $
		\end{enumerate}
		then, the Hamiltonian vector field  has a recursion operator $T_{A}$ given by
		{\small\begin{equation} \label{Eq_3_25}
			T_{A} = \sum_{\mu,\nu =1}^{4}\Bigg(\tilde{M}^{\nu}_{\mu}\dfrac{\partial}{\partial{q^{\nu}}}\otimes dq^{\mu} + \tilde{N}^{\nu}_{\mu}\dfrac{\partial}{\partial{p_{\nu}}}\otimes dp_{\mu} + \tilde{L}^{\nu}_{\mu}\dfrac{\partial}{\partial{q^{\nu}}}\otimes dp_{\mu} +\tilde{R}^{\nu}_{\mu}\dfrac{\partial}{\partial{p_{\nu}}}\otimes dq^{\mu}\Bigg),
			\end{equation}}
		with the corresponding constants of motion 
		{\small\begin{equation*}
			Tr(T_{A}^{h})  = \mathcal{H}^{h} + 2(p_{2}^{h} + p_{3}^{h} + p_{4}^{h}) + \bigg(\dfrac{	\mathcal{H}p_{1}}{p_{1} + v_{s}p_{2}} \bigg)^{h} +  \bigg( \dfrac{ v_{s}p_{2}q^{1}(\mathcal{H} - p_{2}  )}{(p_{1}+ v_{s}p_{2})^{2}} \bigg)^{h}  + (\dot{v}_{s}p_{2}\mathcal{H})^{h},
			\end{equation*}}
		$h \in \mathbb{N},$ where the coordinate dependent quantities $ \tilde{M}^{\nu}_{\mu}, 	\tilde{N}^{\nu}_{\mu}, \tilde{L}^{\nu}_{\mu}$ and  $\tilde{R}^{\nu}_{\mu}$ are expressed as follows:
		{\small\begin{equation*}\label{Eq_s_1}
			\left\{
			\begin{array}{ll}
			\tilde{M}_{1}^{1}= J p_{1}\mathcal{H}_{A} \\
			\tilde{M}_{1}^{2}= p_{2}[J(p_{2} - \mathcal{H}_{A}) - v_{s}] \\
			\tilde{M}_{1}^{k} = Jp_{k}(p_{k} - \mathcal{H}_{A}), \quad k = 3,4 \\
			\tilde{M}_{j}^{j} = p_{j}, \quad j = 2,3;4 \\
			\tilde{M}_{n}^{m} = 0, \quad \mbox{otherwise},
			\end{array}
			\right. ; \quad \left\{
			\begin{array}{ll}
			\tilde{N}_{1}^{1}= \mathcal{H}_{A} \\
			\tilde{N}_{2}^{1}= (p_{2} - \mathcal{H}_{A}) (Jp_{2}- v_{s}) \\
			\tilde{N}_{k}^{1} = Jp_{k}(p_{k} - \mathcal{H}_{A}), \quad k = 3,4 \\
			\tilde{N}_{j}^{j} = p_{j}, \quad j = 2,3;4 \\
			\tilde{N}_{n}^{m} = 0, \quad \mbox{otherwise},
			\end{array}
			\right.
			\end{equation*}
			\begin{equation*}
			\left\{
			\begin{array}{ll}
			\tilde{L}_{1}^{k} = - \tilde{L}_{k}^{1} =  J^{2}p_{k}q^{1}(p_{k} - \mathcal{H}_{A}), \quad k = 2,3,4 \\
			\tilde{L}_{2}^{k} =  J^{2}v_{s}p_{k}q^{1}(\mathcal{H}_{A} - p_{k}  ), \quad k = 2,3,4\\
			\tilde{L}_{j}^{j} = 0, \quad j = 1,3;4 \\
			\tilde{L}_{n}^{m} = 0, \quad \mbox{otherwise},
			\end{array}
			\right. ; \ \left\{
			\begin{array}{ll}
			\tilde{R}_{1}^{1}= \dot{v}_{s}p_{2}\mathcal{H}_{A} \\
			\tilde{R}_{n}^{m} = 0, \quad \mbox{otherwise},
			\end{array}
			\right.
			\end{equation*}
			$n, m = 1,2,3,4$,  $J = \dfrac{1}{p_{1} + v_{s}p_{2}},\ (p_{1} + v_{s}p_{2}) > 0. $ \ }
		
	\end{proposition}
	{\bf Proof.} \\
	Using \eqref{Eq_3_21} and \eqref{Eq_3_21_2}, or \eqref{Eq_s1_2} and \eqref{Eq_s_1_2_2},  it is straightforward to obtain \eqref{Eq_3_25}.
	Furthermore, using conditions $(1)$ and $(2)$, we get
	$ \mathcal{L}_{X_{\mathcal{H}_{A}}}(Tr(T_{A}^{h})) = 0$
	proving that $Tr(T_{A}^{h})$ are constants of motion.
	$\hfill{\square}$

	\item[(ii)] For $ \Delta_{A} = 0$, we obtain the following double root
	$
	\Psi = - v_{s}\alpha_{0}
	$ 
	yielding
	\begin{equation*}
	W = - \alpha_{0} q_{s}  + \alpha_{0} q^{2} + \beta_{0} q^{3} + \gamma_{0} q^{4}, 
	\end{equation*} 
	or, equivalently,
	\begin{equation*} \label{HJE9}
	W = -Q^{2}q_{s}  + \sum_{k=2}^{4} Q^{k}q^{k}
	\end{equation*} 
	in terms of $q^{i}$ and $ Q^{i}$, $ Q^{2} = \alpha_{0}, Q^{3} = \beta_{0},$ and $Q^{4} = \gamma_{0},$ inducing
	the following relationship between the canonical coordinate systems $(Q,P)$ and $(q,p):$
	\begin{equation*}\label{Eq_s2}
	\left\{
	\begin{array}{ll}
	p_{1}= - v_{s}Q^{2} \\
	p_{2} =  Q^{2}     \\
	p_{3} =  Q^{3}  \\
	p_{4} =  Q^{4} 
	\end{array}
	\right.  ; \quad \left\{
	\begin{array}{ll}
	q^{2} = q_{s} - P_{2}  \\
	q^{3} =  - P_{3}  \\
	q^{4} =  - P_{4} 
	\end{array}
	\right. ; \quad
	\left\{
	\begin{array}{ll} 
	P_{2} =   q_{s} - q^{2}\\
	P_{3} = - q^{3} \\
	P_{4} = - q^{4}
	\end{array}
	\right. ; \quad \left\{
	\begin{array}{ll}
	Q^{2} =  p_{2} \\
	Q^{3} =  p_{3} \\
	Q^{4} =  p_{4},
	\end{array}
	\right.
	\end{equation*}
	
	and  the Hamiltonian function $ \mathcal{H}_{A}$
	\begin{equation*} \label{Ham}
	\mathcal{H}_{A} = \dfrac{1}{2} \sum_{k = 2}^{4} (Q^{k})^{2}
	\end{equation*}
	describing the dynamics of a free particle system in the coordinate system $ (Q,P)$, and 
	the associated Hamiltonian vector field 
	\begin{equation*} \label{Hvec}
	X_{\mathcal{H}_{A}} =  - \sum_{k = 2}^{4} Q^{k}\dfrac{\partial}{\partial P_{k}}.
	\end{equation*}
	Since $W$ does not depend on $Q^{1}$ and $P_{1}$, the $(1,1)-$tensor field $T_{A}$ can be given as:
	\begin{equation*} \label{Eq_tensor2}
	T_{A} = \sum_{\nu=2}^{4}Q^{\nu}\Bigg(\dfrac{\partial}{\partial{P_{\nu}}}\otimes dP_{\nu} + \dfrac{\partial}{\partial{Q^{\nu}}}\otimes dQ^{\nu}\Bigg).
	\end{equation*}
	Then, $T_{A}$ satisfies $\mathcal{L}_{X_{\mathcal{H}_{A}}}T_{A} = 0$, $\mathcal{N}_{T_{A}} = 0$ and $deg Q^{\nu} = 2$ proving by Theorem \ref{thi} that  $T_{A}$ is a recursion operator of $X_{\mathcal{H}_{A}}$, with the constants of motion 
	\[
	Tr(T_{A}^{h}) = 2((Q^{2})^{h} + (Q^{3})^{h} + (Q^{4})^{h}), \ h \in \mathbb{N}.
	\]
	In the original coordinate system $(q,p),$ $T_{A}$ becomes 
	\begin{equation*} \label{tens}
	T_{A} = \sum_{\mu,\nu =1}^{4}\Bigg(A^{\nu}_{\mu}\dfrac{\partial}{\partial{q^{\nu}}}\otimes dq^{\mu} + B^{\nu}_{\mu}\dfrac{\partial}{\partial{p_{\nu}}}\otimes dp_{\mu}\Bigg),
	\end{equation*}
	where   
	{\small \[
		A = \left(
		\begin{array}{cccc}
		0 &0 &  0& 0 \\
		\\
		v_{s}p_{2} & p_{2} & 0& 0 \\
		\\
		0  & 0 & p_{3} & 0 \\
		\\
		0& 0 & 0 &  p_{4}
		\end{array}
		\right) \ \mbox{and} \   	B = \left(
		\begin{array}{cccc}
		0 & - v_{s}p_{2} &  0& 0 \\
		\\
		0 & p_{2} & 0& 0 \\
		\\
		0  & 0 & p_{3} & 0 \\
		\\
		0& 0 & 0 &  p_{4}
		\end{array}
		\right),
		\]}
	and the constants of motion  turn to be  
	$Tr(T_{A}^{h})  =   2(p_{2}^{h} + p_{3}^{h} + p_{4}^{h}),\  h \in \mathbb{N}.$
\end{itemize}
\section{Recursion operator of a Hamiltonian vector field in the G\"{o}del metric} \label{sec4}

In this work, as matter of result comparison,  we also consider the G\"{o}del line element $ds^{2}$  in dimensionless cylindrical coordinates \cite{ka}:
\begin{equation*}
ds^{2} = c^{2}dt^{2} - \dfrac{1}{1 + \bigg(\dfrac{r}{2a}\bigg)^{2}} dr^{2} - r^{2} \bigg(1 -  \bigg(\dfrac{r}{2a}\bigg)^{2}\bigg) d\phi^{2} - dz^{2} + \dfrac{2r^{2}c^{2}}{a\sqrt{2}} dtd\phi,
\end{equation*}
where $a$ is a parameter with units of length, which represents
a characteristic distance. In particular, $r=2a$ represents
the critical radius from which CTC can exist [22].

The corresponding tensor metric and its inverse are given, respectivily, by 
{\small\begin{equation*}
	g_{\nu\mu}  = \left(
	\begin{array}{cccc}
	1 & 0 & \dfrac{r^{2}}{a\sqrt{2}} & 0\\
	0 &  - \dfrac{1}{1 + \bigg(\dfrac{r}{2a}\bigg)^{2}} & 0 & 0 \\
	\dfrac{r^{2}}{a\sqrt{2}} & 0 & - r^{2} \bigg(1 -  \bigg(\dfrac{r}{2a}\bigg)^{2}\bigg) & 0 \\
	0 & 0 & 0  & -1\\
	\end{array}
	\right) \ \mbox{and} 
	\end{equation*}
	
	\begin{equation*}
	g^{\nu\mu}  = \left(
	\begin{array}{cccc}
	\dfrac{(2a)^{2} - r^{2}}{(2a)^{2} + r^{2}} & 0 & \dfrac{2a\sqrt{2}}{(2a)^{2} + r^{2}}  & 0\\
	0 &  - \dfrac{(2a)^{2} + r^{2}}{(2a)^{2}}& 0 & 0 \\
	\dfrac{2a\sqrt{2}}{(2a)^{2} + r^{2}} & 0 & -\dfrac{(2a)^{2}}{r^{2}((2a)^{2} + r^{2})} & 0 \\
	0 & 0 & 0  & -1\\
	\end{array}
	\right),
	\end{equation*}}
where we put $ c= 1$.

Now, let  the manifold
$\mathcal{Q} = \mathbb{R}^{4} := \{ q^{1} = t, q^{2} = r, q^{3} = \phi,  q^{4} = z \}, $ where $ t \in (-\infty, + \infty), r \in (0,\infty), \phi \in (0, 2\pi),$ and $ z \in (-\infty, + \infty),$ 
describe the configuration space,
and $\mathcal{T}^{\ast}\mathcal{Q} = \mathcal{Q} \times \mathbb{R}^{4}$ be the cotangent bundle with the local coordinates $(q,p).$  The natural symplectic form and its corresponding Poisson bivector  are given, respectively, by: 
{\small\begin{equation*}\label{Ksy2}
	\omega_{G} = \sum_{\nu = 1}^{4} dp_{\nu} \wedge dq^{\nu}, \ \mathcal{P}_{G} = \sum_{\nu = 1}^{4} \dfrac{\partial}{\partial p_{\nu} }\wedge
	\dfrac{\partial}{\partial q^{\nu}},
	\end{equation*} }
where $\mathcal{T}\mathcal{Q}$
is the tangent bundle.

In the cotangent bundule $\mathcal{T}^{\ast}\mathcal{Q},$  the G\"{o}del metric takes the form:
{\small \begin{equation} \label{god}
	ds^{2} = (cdq^{1})^{2} - \dfrac{1}{1 + \bigg(\dfrac{q^{2}}{2a}\bigg)^{2}} (dq^{2})^{2} - (q^{2})^{2} \bigg(1 -  \bigg(\dfrac{q^{2}}{2a}\bigg)^{2}\bigg) (dq^{3})^{2} - (dq^{4})^{2} + \dfrac{2 (cq^{2})^{2}}{a\sqrt{2}} dq^{1}dq^{3}.
	\end{equation}}
Assuming  $ \dfrac{(q^{2})^{3}}{2a} \ll 1,$ the approximated line element of \eqref{god} is given by 
\begin{equation*} \label{god2}
ds^{2} =  (cdq^{1})^{2} - (dq^{2})^{2} - (q^{2})^{2}(dq^{3})^{2} - 2(q^{2})^{2}\Omega_{G}dq^{1}dq^{2}  - (dq^{4})^{2}+ \mathcal{O} (\Omega_{G}^{2}),
\end{equation*}
where 
$\Omega_{G} = \dfrac{c}{\sqrt{2a}}$ and  $\dfrac{q^{2}}{2a}  \ll 1$.\\
Setting $c = 1$ leads  to 
\begin{equation} \label{god3}
ds^{2} \simeq (dq^{1})^{2} - (dq^{2})^{2} - (q^{2})^{2}(dq^{3})^{2} - 2(q^{2})^{2}\Omega_{G}dq^{1}dq^{2}  - (dq^{4})^{2}+ \mathcal{O} (\Omega_{G}^{2}),
\end{equation}
and the Hamiltonian function
{\small \begin{equation*}
	\mathcal{H}_{G} = \frac{1}{2((q^{2})^{2}\Omega_{G}^{2} +1)} p_{1}^{2} - \dfrac{1}{2}p_{2}^{2} - \frac{1}{2(q^{2})^{2}((q^{2})^{2}\Omega_{G}^{2} +1)} p_{3}^{2} +  \frac{\Omega_{G}}{(q^{2})^{2}\Omega_{G}^{2} +1} p_{1}p_{3} - \dfrac{1}{2}p_{4}^{2}
	\end{equation*}}
with the associated Hamiltonian vector field given by 
\begin{equation*}
X_{\mathcal{H}_{G}} =  \sum_{\mu= 1}^{4} \bigg( U'_{\mu} \dfrac{\partial}{\partial q^{\mu}} - V'_{\mu}\dfrac{\partial}{\partial p_{\mu}}\bigg),
\end{equation*}
where
\begin{align*}
& U'_{1} = \dfrac{1}{(q^{2})^{2}\Omega_{G}^{2} +1} p_{1} + \dfrac{\Omega_{G}}{(q^{2})^{2}\Omega_{G}^{2} +1} p_{3}, \quad U'_{2} = -  p_{2}, \\
&  U'_{3} = \dfrac{\Omega_{G}}{(q^{2})^{2}\Omega_{G}^{2} +1} p_{1} - \dfrac{1}{(q^{2})^{2} ((q^{2})^{2}\Omega_{G}^{2} +1)} p_{3},\quad U'_{4} = - p_{4},\\
& V'_{1} = V'_{2} =V'_{3} = 0, \ \mbox{and} \ V'_{2} = \dfrac{(q^{2})^{2}\Omega_{G}^{2} (2p_{3}^{2} - (q^{2})^{2} p_{1}^{2}) + p_{3} (p_{3} - \Omega_{G}^{3}(q^{2})^{2}p_{1})}{(q^{2})^{3}((q^{2})^{2}\Omega_{G}^{2} +1)^{2}}. 
\end{align*}
The vector field $X_{\mathcal{H}_{G}}$ satisfies the required condition for a Hamiltonian system, $i.e.,$
$\iota_{_{X_{\mathcal{H}_{G}}}} \omega_{G} = -d\mathcal{H}_{G}.
$
Hence, the triplet $(\mathcal{T}^{\ast}\mathcal{Q},\omega_{G} , \mathcal{H}_{G})$ is a Hamiltonian system.

The Hamiltonian-Jacobi equation is given by 
\begin{align*}
E' & = \dfrac{1}{2((q^{2})^{2}\Omega_{G}^{2} +1)} \bigg(\dfrac{\partial W'_{1}}{\partial q^{1}}\bigg)^{2} -\dfrac{1}{2} \bigg(\dfrac{\partial W'_{2}}{\partial q^{2}}\bigg)^{2} - \dfrac{1}{2(q^{2})^{2}((q^{2})^{2}\Omega_{G}^{2} +1)} \bigg(\dfrac{\partial W'_{3}}{\partial q^{3}}\bigg)^{2} \cr
& + \dfrac{\Omega_{G}}{(q^{2})^{2}\Omega_{G}^{2} +1} \dfrac{\partial W'_{1}}{\partial q^{1}}\dfrac{\partial W'_{3}}{\partial q^{3}} - \dfrac{1}{2}\bigg(\dfrac{\partial W'_{4}}{\partial q^{4}}\bigg)^{2},
\end{align*}
where $E'$ is a constant, and $W' = \displaystyle \sum_{\mu = 1}^{4}W'_{\mu} (q^{\mu}) $ is the generating function.

As the Hamiltonian function $\mathcal{H_{G}}$ does not include $q^{1}, q^{2},$ and $q^{3}$, we can set: 
\begin{align*}\label{jac4}
\dfrac{d W'_{1}}{d q^{1}} = \eta' , \   \dfrac{d W'_{3}}{dq^{3}} = \theta', \ \dfrac{d W'_{4}}{d q^{4}} = \vartheta', 
\end{align*}
yielding
{\small \begin{align*}
	E' & = \dfrac{1}{2((q^{2})^{2}\Omega_{G}^{2} +1)} \eta'^{2} -\dfrac{1}{2} \bigg(\dfrac{d W'_{2}}{dq^{2}}\bigg)^{2} - \dfrac{1}{2(q^{2})^{2}((q^{2})^{2}\Omega_{G}^{2} +1)} \theta'^{2}  \cr 
	& + \dfrac{\Omega_{G}}{(q^{2})^{2}\Omega_{G}^{2} +1} \eta'\theta' - \dfrac{1}{2}\vartheta'^{2},
	\end{align*}}
where $ \eta', \theta',$ and $\vartheta'$ are  constants such that the following conditions are satisfied: 
{\small \begin{equation*}
	(i) \ \dfrac{\eta'^{2}}{2E' + \vartheta'^{2}} \ll 1, \ \mbox{with} \ (2E' + \vartheta'^{2}) > 0,
	\ \  (ii) \  \dfrac{(\Omega_{G} \theta')^{2}}{2E' + \vartheta'^{2}} \ll \dfrac{1}{4}, \ \ (iii) \  \dfrac{\eta' \theta'}{2E' + \vartheta'^{2}} \backsimeq \dfrac{1}{2\Omega_{G}^{2}}.
	\end{equation*} }
Thereafter, we get 
\begin{equation} \label{W2}
\bigg(\dfrac{dW'_{2}}{dq^{2}}\bigg)^{2} = \dfrac{1}{(q^{2})^{2}\Omega_{G}^{2} +1} f (q^{2}),
\end{equation}
where  
\begin{equation*}
f (q^{2}) = - (2E' + \vartheta'^{2})\Omega_{G}^{2}(q^{2})^{4} + (- (2E + \vartheta'^{2}) + \eta'^{2} + 2\eta' \theta' \Omega_{G})(q^{2})^{2} - \theta'^{2}.
\end{equation*}
Putting $Z = (q^{2})^{2} $ and considering the above condition $(i)$, $f$ takes the form
\begin{equation*}
f (Z) = - (2E' + \vartheta'^{2})\Omega_{G}^{2}Z^{2}+ (- (2E' + \vartheta'^{2}) + 2\eta' \theta' \Omega_{G})Z - \theta'^{2},
\end{equation*}
with $\Delta_{G} =  (2E'+ \vartheta'^{2}) [ (2E'+ \vartheta'^{2}) - 4\eta'\Phi - 4 \Phi^{2}],$
$\Phi =\theta' \Omega_{G}.$\\  
After computation and using the condition $(ii)$, we obtain  \\ $ \Delta_{G} = 16(2E'+ \vartheta'^{2}) > 0$
affording
\begin{equation*}
Z_{1} = \dfrac{\eta' \theta'}{(2E' + \vartheta'^{2})\Omega_{G}} +  \dfrac{1}{2\Omega_{G}^{2}} \  \mbox{and}  \ Z_{2} = \dfrac{\eta' \theta'}{(2E' + \vartheta'^{2})\Omega_{G}} -  \dfrac{3}{2\Omega_{G}^{2}}.
\end{equation*}
Using the third condition $(iii)$, we get
\begin{align*}
f(q^{2}) = \dfrac{(2E' + \vartheta'^{2})}{\Omega_{G}^{2}} (1 - (q^{2})^{2}\Omega_{G}^{2})((q^{2})^{2}\Omega_{G}^{2} +1), 
\end{align*}
with $(q^{2})^{2}\Omega_{G}^{2} \ll 1$.
Thus, \eqref{W2} becomes

{\small \begin{align*} \label{W21}
	\dfrac{dW'_{2}}{dq^{2}} =  \dfrac{\sqrt{2E' + \vartheta'^{2}}}{\Omega_{G}} \sqrt{(1 - (q^{2})^{2}\Omega_{G}^{2})} \simeq \dfrac{\sqrt{2E' + \vartheta'^{2}}}{\Omega_{G}} \bigg(1 - \dfrac{1}{2}(q^{2})^{2}\Omega_{G}^{2}\bigg), \ W'_{2}(0) = 0
	\end{align*}}
affording
\begin{align*} 
W'_{2} \simeq \dfrac{\sqrt{2E' + \vartheta'^{2}}}{\Omega_{G}}q^{2} \bigg(1 - \dfrac{1}{6}\Omega_{G}^{2}(q^{2})^{3}\bigg) \simeq \dfrac{\sqrt{2E' + \vartheta'^{2}}}{\Omega_{G}}q^{2}.
\end{align*}
Putting $ Q^{1} = E', Q^{2} = \eta',  Q^{3} = \theta' , \ \mbox{and}\ Q^{4} = \vartheta',$ we have
\begin{align*} 
W'\simeq  Q^{2}q^{1} + \dfrac{\sqrt{2Q^{1} + (Q^{4})^{2}}}{\Omega_{G}}q^{2} + Q^{3}q^{3} + Q^{4}q^{4}.
\end{align*}

Then, we obtain the following relationship between the canonical coordinate systems $(Q,P)$ and $(q,p):$
\begin{equation}\label{Eq_s1_4}
\left\{
\begin{array}{ll}
p_{1}= Q^{2} \\
p_{2} =  \dfrac{\sqrt{2Q^{1} + (Q^{4})^{2}}}{\Omega_{G}}     \\
p_{3} =  Q^{3}  \\
p_{4} =  Q^{4} 
\end{array}
\right.  ; \quad \left\{
\begin{array}{ll}
q^{1} = - P_{2}     \\
q^{2} = -P_{1}\Omega_{G}\sqrt{2Q^{1} + (Q^{4})^{2}}  \\
q^{3} =  - P_{3}  \\
q^{4} =  - P_{4} +  Q^{4}P_{1} 
\end{array}
\right.
\end{equation}
\begin{equation}\label{Eq_s_1_2_4}
\left\{
\begin{array}{ll}
P_{1} = -\dfrac{q^{1}}{ \Omega_{G}^{2}p_{2}} \\
P_{2} = - q^{1}\\
P_{3} =   - q^{3} \\
P_{4} = - \dfrac{p_{4}q^{2}}{ \Omega_{G}^{2}p_{2}}  - q^{4}
\end{array}
\right. ; \quad \left\{
\begin{array}{ll}
Q^{1} = \mathcal{H}'_{G}\\
Q^{2} =  p_{1} \\
Q^{3} =  p_{3} \\
Q^{4} =  p_{4}.
\end{array}
\right.
\end{equation}

In terms of the canonical coordinate system $(Q,P)$, the vector field $ X_{\mathcal{H}_{G}}$ and the symplectic form $\omega_{G}$ are written as: 
\[
X_{\mathcal{H}_{G}} = \{\mathcal{H}_{G},.\} = - \dfrac{\partial}{\partial{P_{1}}} ; \quad \omega_{G} = \sum_{\nu = 1}^{4}dP_{\nu}\wedge dQ^{\nu}.
\]
By the Lemma \ref{lem_1},  a $(1,1)-$tenseur field $T_{G}$ can be expressed as: 
\[
T_{G} = \sum_{\nu =1}^{4}Q^{\nu}\bigg(\dfrac{\partial}{\partial{P_{\nu}}} \otimes dP_{\nu} + \dfrac{\partial}{\partial{Q^{\nu}}} \otimes dQ^{\nu}\bigg),
\]
where the constants of motion   are:
\[
Tr(T_{G}^{h}) = 2((Q^{1})^{h} + (Q^{2})^{h} + (Q^{3})^{h} + (Q^{4})^{h}), \ h \in \mathbb{N}.
\]
We arrive at:
\begin{proposition}
	Under the condition   
	\begin{equation} \label{p2}
	V'^{h}_{2} \simeq \bigg(\dfrac{p^{2}_{2}}{q^{2}}\bigg)^{h}, \ h \in \mathbb{N},
	\end{equation}
	the Hamiltonian vector field $X_{\mathcal{H}_{G}}$ in the G\"{o}del metric \eqref{god3} has a recursion operator $T_{G}$ in the original coordinate system $ (q,p)$ given by
	\begin{equation*} \label{Eq_3_27}
	T_{G} = \sum_{\mu,\nu =1}^{4}\Bigg(\tilde{A}^{\nu}_{\mu}\dfrac{\partial}{\partial{q^{\nu}}}\otimes dq^{\mu} + \tilde{B}^{\nu}_{\mu}\dfrac{\partial}{\partial{p_{\nu}}}\otimes dp_{\mu} + \tilde{C}^{\nu}_{\mu}\dfrac{\partial}{\partial{q^{\nu}}}\otimes dp_{\mu} +\tilde{D}^{\nu}_{\mu}\dfrac{\partial}{\partial{p_{\nu}}}\otimes dq^{\mu}\Bigg),
	\end{equation*}
	where	
	\begin{equation*}\label{Eq_s_2}
	\left\{\begin{array}{ll}
	\tilde{A}_{j}^{j}= p_{j} \\
	\tilde{A}_{2}^{2}= \mathcal{H}_{G} \bigg(1 +  q^{2}V'_{2}S\bigg) \\
	\tilde{A}_{2}^{4} = - p_{4}p_{2}S \bigg(\mathcal{H}_{G} + U'_{4} \bigg) \\
	\tilde{A}_{n}^{m} = 0, \quad \mbox{otherwise}
	\end{array}
	\right. ; \quad \left\{
	\begin{array}{ll}
	\tilde{B}_{j}^{j}=p_{j} , \quad j = 1,3,4 \\
	\tilde{B}_{k}^{2}= \mathcal{H}_{G} U'_{k}p_{2}S,\quad k = 1,2,3 \\
	\tilde{B}_{4}^{2} = - p_{4}p_{2}S \bigg(\mathcal{H}_{G} + U'_{4} \bigg)\\
	\tilde{B}_{n}^{m} = 0, \quad \mbox{otherwise}
	\end{array}
	\right.
	\end{equation*}
	\begin{equation*}\label{Eq_s_3}
	\left\{\begin{array}{ll}
	\tilde{C}_{i}^{2}= \mathcal{H}_{G} U'_{i}q^{2}S, \quad i = 1,3 \\
	\tilde{C}_{2}^{2}=  \mathcal{H}_{G} q^{2}\Omega_{G}^{2}S\bigg( - p_{2} + U'^{3}_{2}S \bigg) \\
	\tilde{C}_{4}^{2} =  - p_{4}S\bigg(\mathcal{H}_{G} + U'_{4} \bigg)\\
	\tilde{C}_{2}^{4} =   p_{4}S \bigg(\mathcal{H}_{G} + U'_{4} \bigg)\\
	\tilde{C}_{n}^{m} = 0, \quad \mbox{otherwise},
	\end{array}
	\right. ; \quad \left\{
	\begin{array}{ll}
	\tilde{D}_{2}^{2}= \mathcal{H}_{G} V'_{2}p_{2}S ;\\
	\tilde{D}_{n}^{m} = 0, \quad \mbox{otherwise},
	\end{array}
	\right. ; \ S = \dfrac{1}{\Omega_{G}^{2}p_{2}^{2}},
	\end{equation*}
	$  \Omega_{G}^{2}p_{2}^{2} > 0, n, m = 1,2,3,4.$	
	
	The constants of motion in the original coordinate system $(q,p)$ are $Tr(T_{G}^{h}),\ h \in \mathbb{N}:$
	\begin{align*}
	Tr(T_{G}^{h})  &=  2(p_{2}^{h} + p_{3}^{h} + p_{4}^{h}) + \mathcal{H}_{G}^{h} \bigg(1 +  \dfrac{q^{2}V'_{2}}{\Omega_{G}^{2}p_{2}^{2}}\bigg)^{h} + \bigg(-\dfrac{\mathcal{H}_{G}}{\Omega_{G}^{2}}\bigg)^{h} \cr  & + \bigg(-\dfrac{\mathcal{H}_{G} q^{2}}{p_{2}}\bigg)^{h} \bigg( 1 + \dfrac{1}{\Omega_{G}^{2}} \bigg)^{h} + \bigg(\dfrac{\mathcal{H}_{G} V'_{2}}{\Omega_{G}^{2}p_{2}}\bigg)^{h}.
	\end{align*}
	
\end{proposition}

{\bf Proof.} \\
Using \eqref{Eq_s1_4} and \eqref{Eq_s_1_2_4}, and after some computations, we obtain \eqref{Eq_3_27}.

In addition, from condition \eqref{p2}
{\small\begin{equation*}
	\mathcal{L}_{X_{\mathcal{H}_{G}}} (Tr(T_{G}^{h}) )  = 0.
	\end{equation*}}
Hence, 

{\small$2(p_{2}^{h} + p_{3}^{h} + p_{4}^{h}) + \mathcal{H}_{G}^{h} \bigg(1 +  \dfrac{q^{2}V'_{2}}{\Omega_{G}^{2}p_{2}^{2}}\bigg)^{h} + \bigg(-\dfrac{\mathcal{H}_{G}}{\Omega_{G}^{2}}\bigg)^{h} + \bigg(-\dfrac{\mathcal{H}_{G} q^{2}}{p_{2}}\bigg)^{h} \bigg( 1 + \dfrac{1}{\Omega_{G}^{2}} \bigg)^{h} + \bigg(\dfrac{\mathcal{H}_{G} V'_{2}}{\Omega_{G}^{2}p_{2}}\bigg)^{h}$}

are constants of motion.
$\hfill{\square}$

It is worth noticing that in the 	coordinate system $(Q,P),$  for 	$\Delta_{A} > 0$,
the Alcubierre and 
G\"{o}del metrics  have the same Hamiltonian vector field $X_{H}$ and the same recursion operator $T$, which thus  induce the same dynamics.

\section{Master symmetries}\label{sec5}
Provided the above common dynamical characteristics,  let us consider the Hamiltonian system $(\mathcal{T}^{\ast}\mathcal{Q}, \omega, Q^{1}),$ for which the Hamiltonian function $H$,
the vector field $ X_{0}$, the symplectic form $\omega$, and the bivector field are given in both the  Alcubierre and G\"{o}del metrics by: 
{\small\[
	H = Q^{1}; \quad  X_{0} = \{Q^{1},.\} = - \dfrac{\partial}{\partial{P_{1}}} ; \quad \omega = \sum_{\nu = 1}^{4}dP_{\nu} \wedge dQ^{\nu}; \quad \mathcal{P} = \sum_{\nu = 1}^{4} \dfrac{\partial}{\partial P_{\nu} }\wedge
	\dfrac{\partial}{\partial Q^{\nu}}.
	\]}
Introduce the  vector fields $ Y_{j} \in \mathcal{T}^{\ast}\mathcal{Q},$
{\small\begin{equation*}
	Y_{j} = \sum_{\nu=1}^{4}(Q^{\nu})^{j}\bigg( (j + 1)P_{\nu}\dfrac{\partial}{\partial{P_{\nu}}} - Q^{\nu}\dfrac{\partial}{\partial{Q^{\nu}}}\bigg), \quad j \in \mathbb{N},
	\end{equation*}}
satisfying the relation
{\small\begin{equation*}
	\iota_{_{Y_{j}}}\omega = -d\tilde{H}_{j}, \quad \mbox{with} \quad	\tilde{H}_{j} = - \sum_{\nu = 1}^{4}(Q^{\nu})^{j + 1}P_{\nu}.
	\end{equation*}}
The  symplectic structure $\omega$ generates a set of Hamiltonian systems on the same manifold $\mathcal{T}^{\ast}\mathcal{Q}.$
The Lie bracket  between the vector fields $X_{i}$ and $Y_{j}$ obeys the relations
{\small \begin{equation} \label{sym}
	[X_{i}, Y_{j}] = X_{i + j},\quad [X_{i}, X_{i + j}] = 0, \quad \mbox{with}\quad X_{i + j} = - (j+1)(i+j+1) (Q^{1})^{ i + j  }\dfrac{\partial}{\partial{P_{1}}}, \ i,j \in \mathbb{N}.
	\end{equation}}
This is well illustrated in  Fig $1$.
In differential geometric terms, $Y_{j}$  and 	$\tilde{H}_{j}$ are called  {\it master symmetries} for 	$X_{i}$ and   {\it master integrals}, respectively, \cite{cas,dam,fer2,ra,ra3}.

\begin{figure}[!th]
	\centering
	\includegraphics[width=12cm, height=4cm]{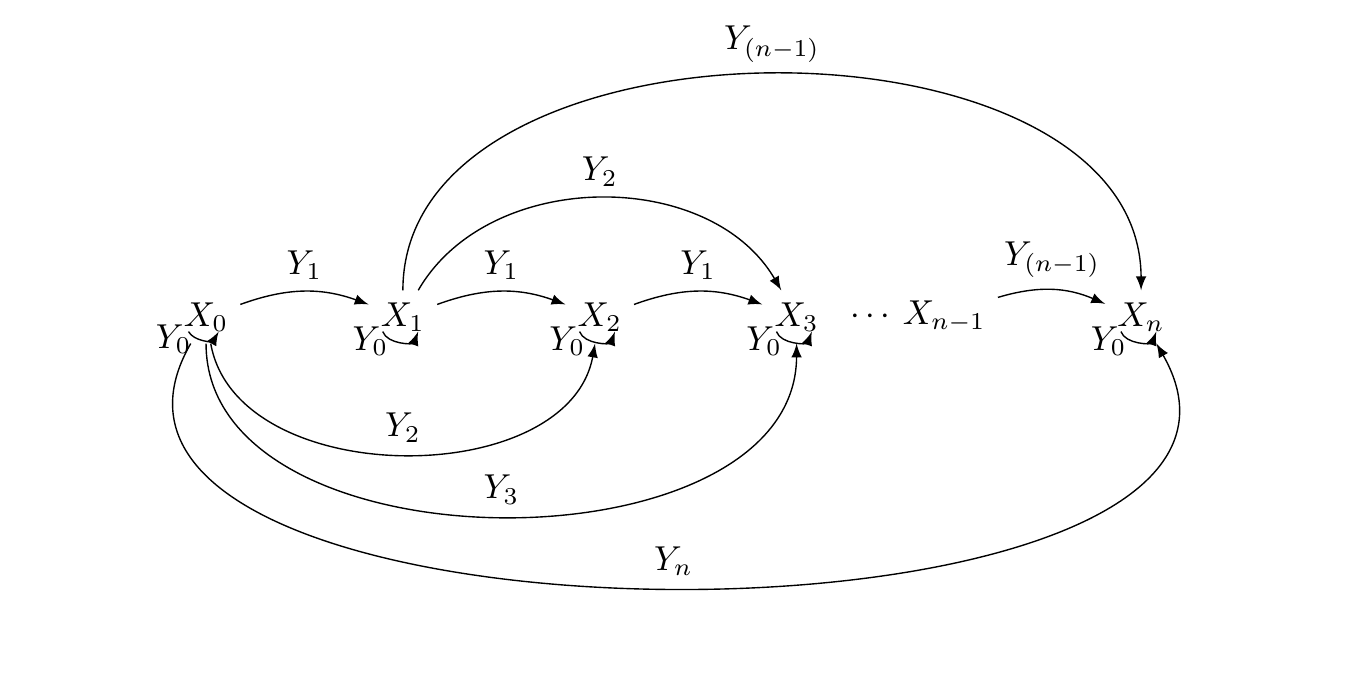}
	\caption{Diagrammatical illustration of equation \eqref{sym}.}
	\label{Graphs2}
\end{figure}
From the master integrals $\tilde{H}_{j}$, we can generate a family  of Hamiltonian functions:
\begin{equation*} \label{symv2}
H_{i +j} := \{H_{i}, \tilde{H}_{j}\} = (i +1) (Q^{1})^{i +j + 1}, \ \mbox{with} \ H_{0} = H, \ i, j \in \mathbb{N}.	
\end{equation*}
The recursion operator $T,$
\[
T = \sum_{\nu =1}^{4}Q^{\nu}\bigg(\dfrac{\partial}{\partial{P_{\nu}}} \otimes dP_{\nu} + \dfrac{\partial}{\partial{Q^{\nu}}} \otimes dQ^{\nu}\bigg),
\]
can be written as:
\[
T = \mathcal{P}_{1}\circ \mathcal{P}^{-1},
\]
where 
\begin{equation*}
\mathcal{P}_{1} =  \sum_{\nu = 1}^{4} Q^{\nu} \dfrac{\partial}{\partial P_{\nu} }\wedge
\dfrac{\partial}{\partial Q^{\nu}}
\end{equation*}
and $\mathcal{P}$ are two compatible Poisson bivectors  
with vanishing Schouten-Nijenhuis bracket 
$[\mathcal{P}, \mathcal{P}_{1}]_{NS} = 0.$

Now, introducing the following Poisson bracket $\{.,.\}_{1}$
\begin{equation*}
\{f,g\}_{1} := \sum_{\nu = 1}^{4} Q^{\nu}\Bigg(\dfrac{\partial{f}}{\partial{P_{\nu}}}\dfrac{\partial{g}}{\partial{Q^{\nu}}} - \dfrac{\partial{f}}{\partial{Q^{\nu}}}\dfrac{\partial{g}}{\partial{P_{\nu}}}\Bigg),
\end{equation*}
with respect to the symplectic form $ \omega_{1} = \displaystyle\sum_{\nu = 1}^{4} (Q^{\nu})^{-1} dP_{\nu} \wedge dQ^{\nu},$
we get

$	
X_{i} = \{\bar{H}_{i},.\} = \{\bar{H}_{i + 1},.\}_{1}, \ \bar{H}_{0} = H, \ \bar{H}_{1} = \ln(Q^{1}), \ \bar{H}_{j}= - \dfrac{1}{j(Q^{1})^{j}}, \ Q^{1} \neq 0,$  $  X_{0} = - \dfrac{\partial{}}{\partial{P_{1}}}, \ X_{1} = - \dfrac{1}{Q^{1}}\dfrac{\partial{}}{\partial{P_{1}}}, \  X_{j} = - \dfrac{1}{(Q^{1})^{j +1}}\dfrac{\partial{}}{\partial{P_{1}}}, \ j= 2,3,...,n ; \ n,i  \in \mathbb{N},$
proving that $X_{i}$ are bi-Hamiltonian vector fields defined by the two Poisson bivectors $\mathcal{P}$ and $\mathcal{P}_{1}.$
Then, the quadruple $(\mathcal{Q},\mathcal{P},\mathcal{P}_{1},X_{i} )$ is a bi-Hamiltonian system for each $i$.

In addition, we have
{\small\begin{eqnarray*}
		& &\mathcal{L}_{Y_{0}} (\mathcal{P}) = 0, \ (\tilde{\alpha} = 0), \quad \mathcal{L}_{Y_{0}} (\mathcal{P}_{1}) = - \sum_{\nu = 1}^{4} Q^{\nu} \dfrac{\partial}{\partial P_{\nu} }\wedge
		\dfrac{\partial}{\partial Q^{\nu}}, \ (\tilde{\beta} = -1), \cr
		& &  \mathcal{L}_{Y_{0}} (H) = - Q^{1} = -H,\ (\tilde{\gamma} = - 1).
\end{eqnarray*}}
We conclude that the vector field

\begin{equation*}
Y_{0} = \sum_{\nu=1}^{4}\bigg( P_{\nu}\dfrac{\partial}{\partial{P_{\nu}}} - Q^{\nu}\dfrac{\partial}{\partial{Q^{\nu}}}\bigg), 
\end{equation*}
is conformal symmetry for  $\mathcal{P}, \mathcal{P}_{1}$ and $H$ \cite{fer2}.
Defining now the families of quantities $X'_{h}, \ Y'_{h}, \ \mathcal{P}'_{h}, \ \omega'_{h} $ and $dH'_{h}$  by
{\small\begin{equation*}
	X'_{h}:= T^{h}X_{0}, \ \mathcal{P}'_{h}:= T^{h}\mathcal{P}, \ \omega'_{h}:= (T^{\ast})^{h}\omega', \ Y_{h}:= T^{h} Y_{0}, \ dH'_{h} :=(T^{\ast})^{h} dH, \quad h \in \mathbb{N},
	\end{equation*}}
where   $T^{\ast} :=  \mathcal{P}^{-1}\circ \mathcal{P}_{1} $ is the adjoint of $T,$
we get
{\small \begin{align*}
	& \mathcal{P}'_{h} = \sum_{\nu = 1}^{4} (Q^{\nu})^{h} \dfrac{\partial}{\partial P_{\nu} }\wedge
	\dfrac{\partial}{\partial Q^{\nu}}, \  Y'_{h} =  \sum_{\nu=1}^{4}(Q^{\nu})^{h}\bigg( P_{\nu}\dfrac{\partial}{\partial{P_{\nu}}} - Q^{\nu}\dfrac{\partial}{\partial{Q^{\nu}}}\bigg), \  X'_{h} = - (Q^{1})^{h}\dfrac{\partial}{\partial{P_{1}}}, \cr
	& \omega'_{h} = \sum_{\nu = 1}^{4}(Q^{\nu})^{h}dP_{\nu} \wedge dQ^{\nu} ,\ dH'_{h} = (Q^{1})^{h}dQ^{1} \ \mbox{and}  \ H'_{h} = \dfrac{1}{h + 1}(Q^{1})^{h + 1}
	\end{align*}}
leading to the following plethora of conserved quantities:
{\small\begin{align*}
	&\mathcal{L}_{Y'_{h}} (Y'_{l}) = (h -l)Y'_{l +h}, \ \mathcal{L}_{Y'_{h}} (X'_{l}) = - (l + 1)X'_{l +h}, \ \mathcal{L}_{Y'_{h}} (\mathcal{P}'_{l}) =  (h - l )\mathcal{P}'_{l + h},  \cr
	& \mathcal{L}_{Y'_{h}} (\omega'_{l}) = -(l + h )\omega'_{l + h}, \ \mathcal{L}_{Y'_{h}} (T) = - T^{1 + h}, \ \langle dH'_{l} , Y'_{h} \rangle  = -(h + l+ 1)H'_{l + h}, \ l \in \mathbb{N}
	\end{align*}}
satisfying
{\small\begin{align*}
	&\mathcal{L}_{Y'_{h}} (Y'_{l}) = (\tilde{\beta} - \tilde{\alpha})(l - h)Y'_{(l +h)}, \  \mathcal{L}_{Y'_{h}} (X'_{l}) = (\tilde{\beta} + \tilde{\gamma} + (l - 1)(\tilde{\gamma} -\tilde{\alpha})) X'_{l + h}, \cr &\mathcal{L}_{Y'_{h}} (\mathcal{P}'_{l}) = (\tilde{\beta} + (l -h -1)(\tilde{\beta} - \tilde{\alpha})) \mathcal{P}'_{l + h}, \ \mathcal{L}_{Y'_{h}}(\omega'_{l})=  (\tilde{\beta} + (l + h - 1)(\tilde{\beta} - \tilde{\alpha})) \omega'_{l + h}, \cr
	& \mathcal{L}_{Y'_{h}}(T) = (\tilde{\beta} - \tilde{\alpha}) T^{1 + h} , \ \langle dH'_{l} , Y'_{h} \rangle  = 
	(\tilde{\gamma} + (l + h)(\tilde{\beta} - \tilde{\alpha}))H'_{l + h},
	\end{align*}}
analogue to the  Oevel formulae (see \cite{oev,fer2,smir,smir2}).

\section{Conclusion} \label{sec6}
In this paper, we have analyzed in detail the dynamics of  a spaceship in Alcubierre and G\"{o}del metrics.
We have derived the Hamiltonian vector fields governing the system evolution,
constructed and discussed related recursion operators generating the constants of motion. Besides, we have proved the existence of a bi-Hamiltonian structure in the considered canonical coordinate system and 
computed  conserved quantities using  the corresponding master symmetries. 

This study has shown that Hamiltonian dynamics hints at a connection between the geometry
of the physical system  and conservation laws using the Poisson bracket. Our physical systems in Alcubierre and G\"{o}del metrics are symplectic manifolds equipped with  Hamiltonian vector fields.  In this connection, the spaceship positions  on the manifolds are viewed as states and vector fields as laws governing how those states evolve.

We have observed that the spaceship obeys the same dynamics for particular choices of the Alcubierre and G\"{o}del metrics. Indeed, using appropriate parametrizations, the Hamiltonian vector fields and the recursion operators have been expressed in  identical way for both the metrics. The only difference between them has been  the relationship between the original coordinates and the new coordinates.  Further, we have noticed that the Hamiltonian function of the spaceship remains constant along the trajectories   (also called integral curves) for Hamiltonian vector fields.

We have used the recursion operator to compute the constants of motion, i.e., first integrals, which are an important step in the study of the dynamics of the spaceship. Each Hamiltonian vector field $X_H$ is its own first integral, $X_H(H):= \{H,H\}= 0$ due to the anti-symmetry of the Poisson bracket. This is characteristic of the physical principle of energy conservation.

Finally, from this study, we infer the formulation of a generalized Poisson bracket as follows: 
\begin{equation*}
\{f,g\}_{j} := \sum_{\nu = 1}^{4} (Q^{\nu})^{j}\Bigg(\dfrac{\partial{f}}{\partial{P_{\nu}}}\dfrac{\partial{g}}{\partial{Q^{\nu}}} - \dfrac{\partial{f}}{\partial{Q^{\nu}}}\dfrac{\partial{g}}{\partial{P_{\nu}}}\Bigg), \ j\in \mathbb{N}
\end{equation*}
yielding a set of bi-Hamiltonian vector fields
\begin{equation*}
X_{i} = \{\bar{H}_{i},.\} = \{\bar{H}_{i + j},.\}_{j}, \quad i , j \in \mathbb{N},
\end{equation*}
which can allow a straightforward extension of all previous results. 

\section*{Conflict of Interest}
The authors declare that they have
no conflicts of interest.

\subsection*{Acknowledgments}
We thank the referee and the Editorial Board for their useful comments, which permit to improve the paper.
The ICMPA-UNESCO Chair is in partnership 
with the Association pour la Promotion Scientifique de l'Afrique
(APSA), France, and Daniel Iagolnitzer Foundation (DIF), France,
supporting the development of mathematical physics in Africa.
M. M. is supported by the Faculty of Mechanical Engineering,
University of Ni\v s, Serbia, Grant ``Research and development of
new generation machine systems in the function of the technological
development of Serbia''.

\end{document}